# Overview of contemporary systems driven by open-design movement

Aditya M. Medhi[1], Abhishek D. Patange[1*], Sujit S. Pardeshi[1], R. Jegadeeshwaran[2], Mustafa Kuntoğlu[3]

[1]Department of Mechanical Engineering, College of Engineering Pune, Pune – 411005, India

[2]School of Mechanical Engineering, Vellore Institute of Technology, Chennai, – 600127, India

[3]Department of Mechanical Engineering (Faculty of Technology), Selcuk University, Konya, Turkey

*Corresponding Author: Abhishek D. Patange,*

*Email ID:* abhipatange93@gmail.com / adp.mech@coep.ac.in

**Abstract:**

The movement for open-design focuses on the creation of machines, physical systems, and products using design information shared publicly. It consists of the development of systems incorporating open-source hardware and software which can be easily/freely customized and implemented. Generally, this movement is adopted through the Internet and usually executed without economic recompense. The aim and idea of this movement is similar to the open-source movement, however is employed for designing & developing physical systems instead of software system alone. This design necessitates co-creating the end product, which is expected to be designed by the users, in place of an outdoor investor for example a private business. In tune with this, the comprehensive review is carried out wherein a variety of contemporary systems driven by open-design movement for diverse applications is discussed.

## 1. Introduction

The distribution of industrial expertise is found to have its roots in the 18$^{th}$ and 19$^{th}$ century [1, 2]. Rigorous coining of research has scotched that period of widespread sharing of knowledge [3]. In recent times, tenets of such kind of design are found to be linked to free and open-source movements of this software system [4]. In 1997 Tim O'Reilly, Larry Augustin and Eric S. Raymond, founded this term as an alternate illustration of free software [5-7]. Later that year, Bruce Perens issued its formal definition [8]. In the following year, Dr. Sepehr Kiani understood that designers and researchers may possibly gain from such policies, so he persuaded Dr. Ryan Vallance and Dr. Samir Nayfeh of its possible advantages in mechanical design applications [9]. Both of them collectively set up an Open Design Foundation, a non-income conglomerate, and started to develop a definition for open design [9]. This design concept was then used by various other individuals and groups. The doctrines of this design are most nearly akin to that of hardware design which is open sourced, and which came into existence in March of 1998 when Reinoud Lamberts, an alumnus of Technical University of Delft offered his open design circuits' website for the formation of an organization dealing with hardware design [10]. Ronen Kadushin invented the name 'Open Design' while working on his master's dissertation, which was later recognized in the open design manifesto [11]. This movement presently binds two styles. On one side, it provides a support for expanding sophisticated technologies and ventures which could be outside the scope of any country/company and further involves individuals who, devoid of the copy left process, may not work in partnership else. And on the other of it, individuals employ their own time and skills on assignments for the commonweal, conceivably where commercial interest or funding in developing countries is found to be deficient to help them spread economic and ecological technologies. Now a third emerging trend wherein above methods are binded to implement a sophisticated open-source technology [12]. This design possesses a large caliber in shaping future innovation as is seen in the recent research which demonstrates that stakeholders who have collaborated generate innovative and creative designs than those who consult consumers by conventional means [13-20]. Figure 1 shows journey of open source software and hardware from 1991 to today and today to future [114].

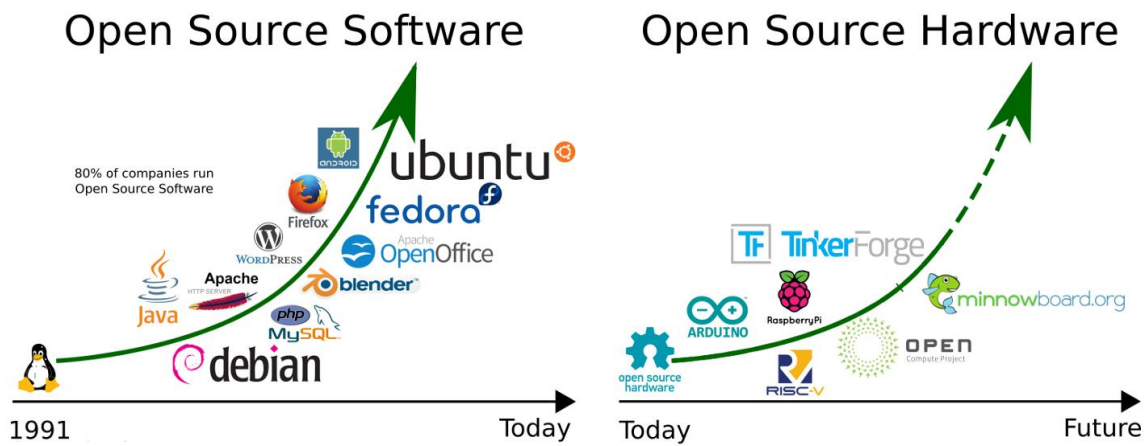

**Figure 1:** Open source software and hardware – from 1991 to today and today to future [114]

## 2. Open source software

Open software is a kind of software which releases a source code using a permit in which the grant is given to the consumers to study, change, distribute and use the software by the patent holder for any purpose. This software can be created in a collective public manner. It is a renowned illustration of open and free partnership. It is easier to obtain instead of commercial software which results in its increased use. Moreover, it has also built loyalty among developers as more and more developers now feel emboldened and get a sense of proprietorship of the final produce [21-25]. Some examples of open-sourced software are:

- Netscape Navigator, which was foundational in the development of famous Mozilla and Mozilla Firefox web browsers.
- StarOffice, which is the basis of LibreOffice and OpenOffice.org office suite.

- SAP DB, which became MaxDB, is now being owned and distributed by MySQL ABInterBase database, which was sourced open by Borland in 2000 is presently existing as an open-source fork (Firebird) and a commercial product.

In detail information of different software is provided here.

- Python: This high-level, general-purpose and interpreted, programming language stresses understandability using its distinguished use of important indentation. Clear and logical code can be written for various small- and large-scale projects because of its language constructs and object-oriented approach. Its garbage-collected and dynamically-typed structure assists numerous programming examples, which includes functional programming and structured object-oriented. It has a comprehensive standard library due to which is known as a "batteries included" language [26].

- Raspbian: It is a Debian-based operating system for the device Raspberry Pi. It is extremely enhanced for single-board computers which are compact and equipped with ARM CPUs. It works properly on all Raspberry Pi devices except for the Pico Controller. Its desktop environment uses a modified LXDE along with Openbox stacking window manager which houses a unique theme. The software includes a version of Wolfram Mathematica, which is an algebra program and a Pi edition of Minecraft and a lightweight version of Chromium web browser [27].

- Keil C Software: Founded in 1982 by Reinhard and Keil Günter, Keil C is German software which is a subordinate of Arm Holdings. It executed the first C compiler which was explicitly designed for 8051 microcontrollers. The broad range of tools which can be used with this software are as follows: ANSI C compiler, macro assemblers, debuggers and simulators, linkers, IDE, library managers, real-time operating systems, and assessment boards for Intel MCS-251,Intel 8051, XC16x/C16x/ST10 families and ARM [28]. Figure 2 shows a comparison between open source software and hardware [115].

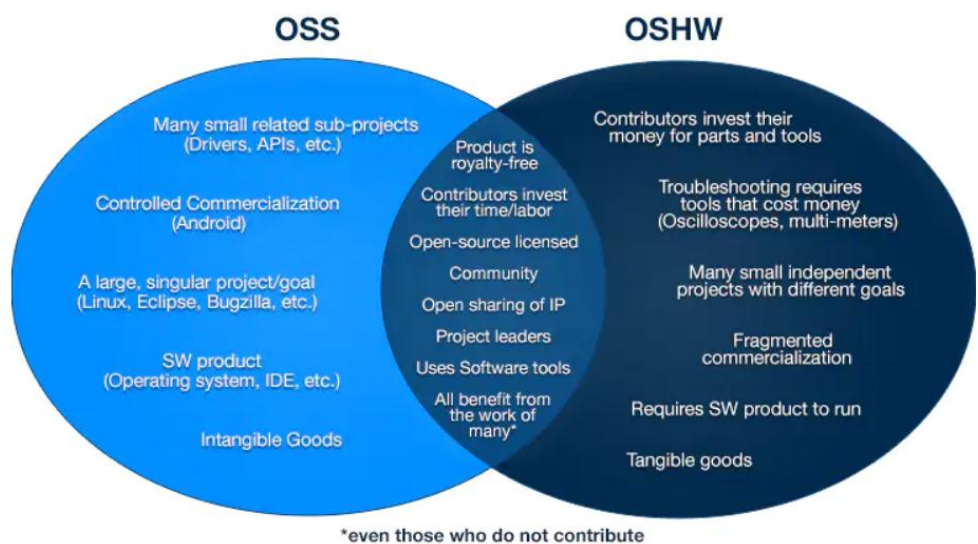

**Figure 2:** A comparison between open source software and hardware [115]

- Visual C++: It is a compiler developed by Microsoft for compiling codes written in various languages such as C++, C++/CLI and C. This trademarked software was initially separate merchandise which later became a portion of Visual Studio and was made accessible in both freeware and trialware systems. Most of the applications need redistributable Visual C++ runtime library packages so that they function accurately. These packages are frequently independently installed which allow manifold applications to use the software requiring only one-time install. They are mainly installed for basic libraries that most functions use [29].
- Apache OpenOffice: It is an open-source office productivity software suite which is one of the beneficiary projects of IBM Lotus Symphony and OpenOffice.org . It is a near companion of Neo-Office and LibreOffice. It houses a spreadsheet, a word processor, a database management application, a formula editor, a drawing application, and a presentation application. It is primarily built for macOS, Windows, and Linux, with ports to other operating systems [30].
- Sun Tracking Software: This device orients payloads, which are generally parabolic troughs, solar panels, Fresnel troughs, lenses, or heliostat mirrors, towards the Sun. For flat-panel photovoltaic systems, trackers minimize the cosine error, which is the angle of incidence between a photovoltaic pane and incoming sunlight. As this angle decreases, the amount of energy produced increases. Trackers are used to allow the optical components in concentrated solar power (CSP) and concentrator photovoltaic (CPV)) applications. The optics must be oriented suitably to amass energy with high efficiency [31].
- Free scale CodeWarrior: Developed on the Macintosh by Metrowerks , It provides an integrated development environment (IDE) for compiling, debugging and editing the software for many microprocessors and microcontrollers and digital signal controllers so as to be used in embedded systems. CodeWarrior rapidly developed into the de facto standard system for development for Mac during Apple's shift to the PPC. It went on to become a major part of the software stack for Motorola's diverse microcontroller lines which ultimately steered them to purchase Metrowerks in 1999 [32].
- VSM Software: Value Stream mapping software is software which supports the preparation of analyzing value stream maps. This software designs maps by making use of a series of symbols that represent activities and material/information flow, which further acts as a supplement to hand calculations. These maps are not ardent to build and utilization of the software could assist in speeding up of the process which can further abridge calculations [33].
- Zimbra: Previously well-known as the Zimbra Collaboration Suite, Zimbra Collaboration is a collective software suite that comprises a web client and an email server. Two editions of this software are offered: a commercially endorsed and an open source version. It includes closed-source parts like exclusive Messaging Application Programming Interface connector for Outlook for contact and calendar harmonization. It utilizes many open-source ventures. It reveals a SOAP application programming interface to each of its functionality. This server can work on most distributions of Linux and also on a Windows Server, by making use of a container technology and a computer-generated machine . For messaging, it bolsters CardDAV, CalDAV and SMTP, LDAP for Microsoft Active Directory, and directory services. It utilizes MTA functionality by using Postfix. For email encryption and signing, it consists of technology from SpamAssassin, DSPAM and ClamAV for anti-malware elements. Zimbra can orchestrate contacts, calendar, and mail and items using open-source mail consumers like Evolution and Mozilla Thunderbird. It can also synchronize with patented customers like Apple Mail and Microsoft Outlook, either by using exclusive connections or by using the ActiveSync protocol Along with this; it also offers local two-way sync to most mobile gadgets [34].

- SugarCRM: It is an open-sourced a customer relationship management software corporation situated in Cupertino city of California. It generates the web application Sugar whose functions include marketing campaigns, sales-force automation, collaboration, customer support Social CRM, reporting, and Mobile CRM [35].

**3. Open source hardware**

Open-source hardware (OSH) is a hardware whose model is distributed in real time so that everyone can research, alter, dispense, make, and advertise it. It comprises of physical objects of technology which are offered and designed under this open movement for the benefit of technical enthusiasts. This helps in easing the concern of exclusive device drivers for the open and free software association, yet, it is not obligatory for it, and must not be jumbled with the idea of open and free source documentation for recorded hardware. The word hardware in this concept is used in rival with the term software of the same concept from historical times. Nevertheless, as more hardware products, which are not electronic, are made open sourced, this term finds it use in a broader sense which is that of a physical product. Figure 3 shows the forms of openness in open-source hardware [118].

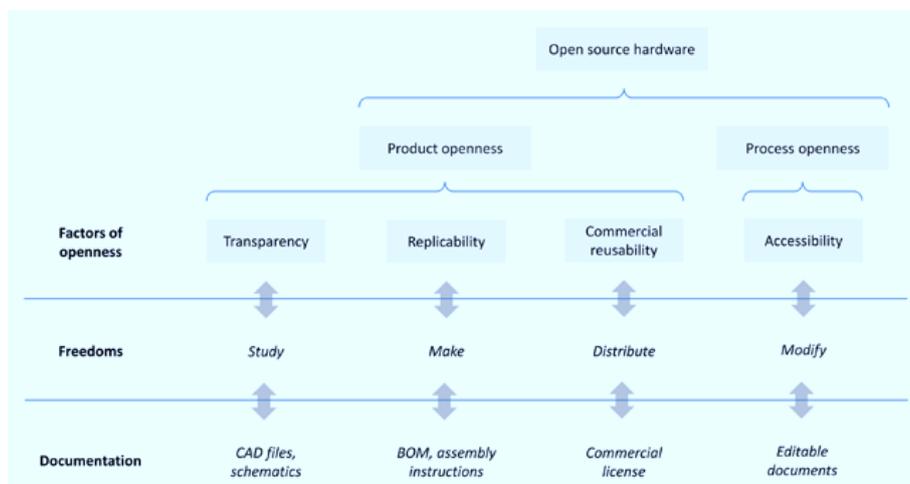

**Figure 3:** Forms of openness [118]

OSH is an expansion of the merchandise enhancement and IP management methods from open-source software (OSS) into tangible entities. In this evolving discipline, free release and transparency of artifact-associated data aims to encourage democratic, participative, society-based types of product development in which any fascinated individual can get intertwined, regardless of his administrative or geological environment. This field has gone ahead of electronic hardware to encompass a wider range of product classes such as vehicles, medical equipment's, and machine tools. In this sense, hardware means any type of quantifiable product be it mechanical, electronic etc. An example of this kind of hardware includes OpenBeam, Wikihouse and Hovalin [36-40]. R. Mies et al. [41] aspires the readers to the aiding groupware for Open-Source Product Development (OSPD) by uncovering its constraints and offering a standard of solutions presently used by practitioners of OSPD. It is exemplified by two kinds of openness: product openness where the "source" of the hardware is made freely accessible and process openness where any concerned individual can take involvement in the process of development [42]. Figure 4 shows an open hardware designing with IoT– mangOH [116].

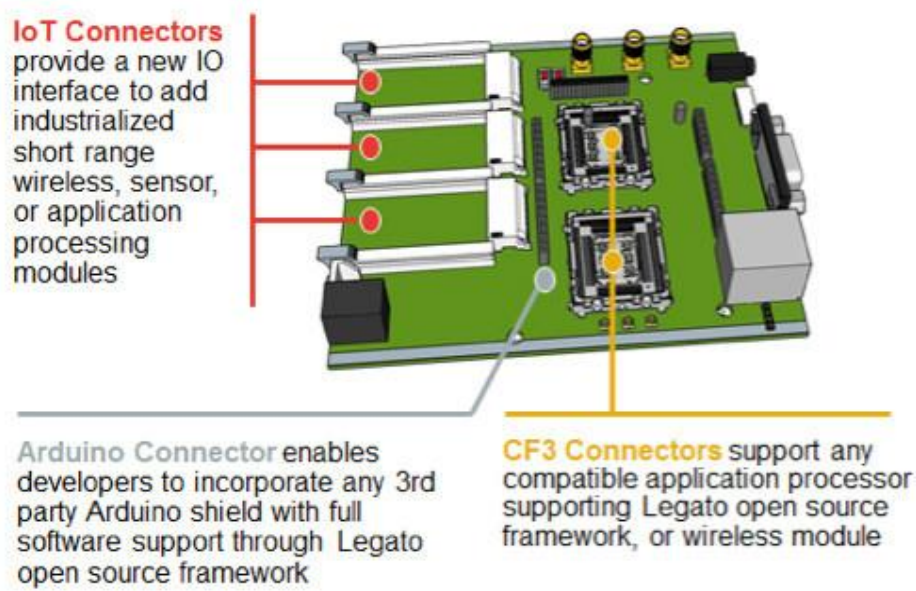

**Figure 4:** An open hardware designing with IoT– mangOH [116]

According to Open Source Hardware Association [43], the phrase Open-Source Hardware belongs to a product whose design has been announced to the community such that everyone can get, alter, supply, and make use of it. An OSH product is a natural product whose credentials are issued under a license allowing everyone with distribution and production rights and is comprehensive enough to empower everyone to research and build it beyond. Instead of being of a dual value, openness stretches along a range between two intense states: one is fully closed and the other fully open. On these lines, and as a support to put practical norms in OSH, the scientific article proposed by J. Bonvoisin and R. Mies [44] a straightforward assessment system – called as the Open-o-Meter. Open-o-Meter (OoM) is a free and open rating method based on an aggregate point scale. It appraises products fairly to eight binary conditions focusing on elements of product and process openness. Based on these eight points, a product is evaluated for its openness. Some of the projects which were evaluated using OoM to check their openness are briefed here. The echOpen is a development targeted at designing a functional low cost and open-source echo-stethoscope. Its process openness is three and product openness is five. This causes in an OoM rating of eight [45]. The inMoov is a completely efficient design of a 3D printed life-size robot. Its process openness is zero and product openness is three. This means that its OoM rating is three [46]. The POM is mentioned to as the planet's first open-source bulk market automobile stage and it is a joint project of Open Motors and Renault. None of the OoM criteria could be fulfilled by the obtainable credentials, causing in a zero OoM value [47]. This meter offers the community with a straightforward list to understand the attempts made by an originator of product to conform to the open-source principles. It also delivers specialists with a strong recommendation to oversee the product information alongside and after the development method. It has grown to be a helpful manual for businesses to propose the incorporation of open-source methodologies in their corporate models [48]. It also exposes the circumstantial and multifactorial environment of openness and offers additional normalization endeavor with a foundation to debate which of these considerations may be deemed compulsory or voluntary. Figure 5 shows some examples of OSHW projects [118].

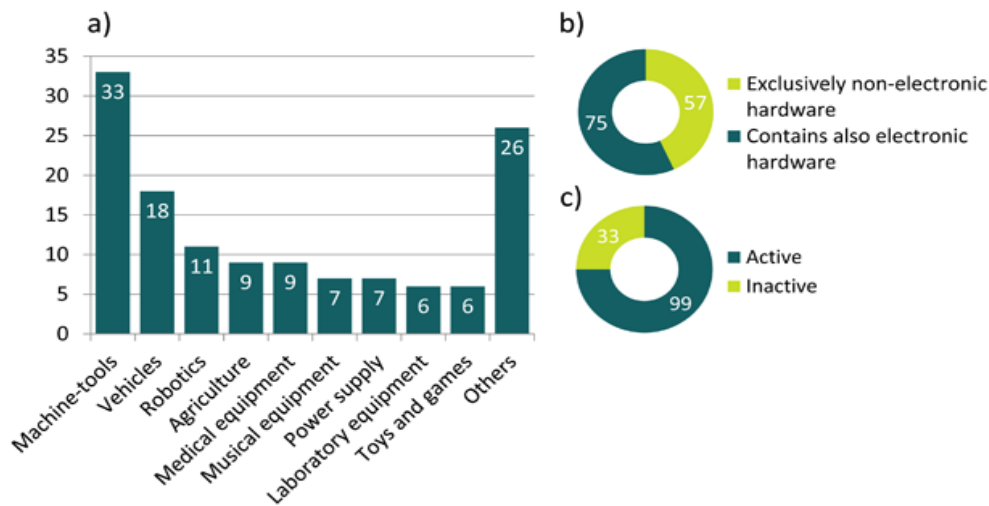

**Figure 5:** Some examples of OSHW projects [118]

## 4. Initiatives related to open design movement

This design is now a neophyte program which consists of many loosely or unrelated connected schemes [49]. Most of them are lone funded ventures, while few bodies have now shifted their focus on under-developed area. In some cases, groups are making a constant attempt to generate a central design which is open and free sourced since it allows for innovation and creativity [50]. Famous organizations based on this concept include:

- Agua Clara: It is an industrial centered project team within Cornell University's Environmental and Civil School of Engineering, which uses open-source technology to design a sustainable water treatment plant. The project's sole objective is to protect and uphold every citizen's fundamental and universal right- access to potable water [51].
- Arduino: Arduino is an open-source electronics platform based on easy-to-use hardware and software. Due to its user-friendly approach, it is now being used as an open-sourced hardware for many engineering projects and applications. It is compatible with any operating system adding to its flexibility. This has proved to be advantageous because of its simple, clear programming environment. Further, it is an economical and open-sourced device which is accessible to all [52].
- Instructables: It is a website having its expertise in DIY projects and is now being owned and operated by Autodesk. It was started by Saul Griffith and Eric Wilhelm in 2005. It collaborates with its members to construct a range of projects. It also features step-by-step instructions for the project followed by visual aids which gives a detailed insight in the projects. It also has a forum in which the enthusiasts can seek immediate answers to their questions [53].
- One Laptop Per Child: It is an initiative started at Massachusetts Institute of Technology in 2005 to give every child in emerging territories a laptop with software and hardware which is open-sourced. This has helped in creating the interest among budding individuals about engineering applications as they are now able to design and learn from it [54].
- Open Cores: It is the most renowned online society for the advancement of gateway IP Cores. It is also a place where such cores are promoted and shared with the inspiration of Open and Free sourced alliance. It presents the source code for various digital gateway assignments and favours the public by giving them a stage where they can list, present, and manage these projects. Moreover, it provides an environment where digital designers meet to promote talk and showcase, their work, passion and creativity. This is carried out using forums, and news collectors [55].
- Open Design Alliance : CAD software It is a not-for-profit corporation which creates SDKs for most of the engineering applications. It also proposes interoperability tools for .dxf, .dgn, .dwg, Autodesk Navisworks, Autodesk Revit, and .ifc files and further provides a technology stack for modeling, visualization, web development, 3D PDF publishing, etcetera [56].
- Open Hardware and Design Alliance: This alliance focuses at promoting the allocation of open designs and hardware. The main idea of this project is an online open service where makers of open designs and hardware register their manufactured goods with a common brand name. This brand name maps four freedoms of open software to physical devices and its documentation. It is very similar to a non-registered brand for hardware and can be contrasted to other certificates such as CE or U.S. Federal Communications Commission mark. OHANDA thus has the role of a self-organized registration authority [57].

- Open Structures: It is an open and a free modular construction model centered on a common geometric grid, known as the OS grid. It was created by Thomas Lommée, a designer by profession, and first exhibited it at the Z33, which houses modern-day art. This tool discovers the likelihood of a modular system wherein everyone designs for each other. Currently, it is building a database in which anybody can share their designs which in turn will be accessible for download by the public. Every component of the design in this system will showcase formerly constructed OS components which were used to create it. Moreover, each part will showcase different component designs which could be created from it [58].
- Open-Source Ecology: Its vision is to create a world of collaborative designs for a inclusive and a transparent economy of abundance. A libre economy, open source, is an economy with high efficiency which is trying to increase innovation and creations by allowing open and free alliances. To achieve this feat, it is presently creating a pack of open and free blueprints for the Global Village Construction Set, which is a pack of the 50 highly significant systems that is required for a modern life to exist. In this process, this organization intends to develop a scalable, modular stage for developing and documenting libre hardware and open source. The present workable execution of this conglomerate is a life size LEGO set of self-replicating and powerful manufacturing devices for dispersed production [59].
- Sensorica: It is an open and free valued network, which is founded in 2011 in Montreal city of Canada, for the purpose of hardware development which is open and free to all. It is a pilot project for commons-centered fellow production which is applied to hardware and is created for the purpose of large-scale operation. It uses the Events, Agents, and Resources, , by considering model as its ground for network resource contribution and planning. It is an Enterprise Resource Planning kind of software featured on the REA model so as to support the intricacy of operations. It collects, interprets, and stores, data from various kinds of movements in the network and binds them to particular resources, agents, and events, so that it keeps a track of the added value on reserve level [60].
- Thingiverse: It is a website whose sole aim is to the share digital design files which are created by the users. Open and free user designs are then sanctioned under the Creative Commons licenses or the GNU General Public License, depending on the user wish. Laser cutters, 3D printers, and production machines and many such machines can be created using this website. It is widely used for DIY products. Several technological designs use this website as a warehouse for pooled dissemination and innovation of source materials to the public [61].
- VIA OpenBook: It is a laptop design from VIA Technologies, which was announced in 2008. Its design of laptop case was published as an open source [62].
- Zoetrope: It is an open and free, vertical axis wind turbine which is manufactured using common engineering materials [63].

- OPENNEXT is a project that enables small and medium-sized enterprises (SMEs) across Europe to engage in communities with consumers and makers in order to fundamentally change how products are designed, produced, and distributed. The project is coordinated by Professor Roland Jochem from the Technische Universität Berlin and brings together a network of partners from 19 EU countries, bridging the gap between business and consumer. Participating companies will openly share ideas and knowledge on digital platforms. The researchers explain that, "we want to empower both companies and consumers by giving them equal access to knowledge in order to co-design and manufacture user-centric products" [64]. Figure 6 is a concept by OPENNEXT.

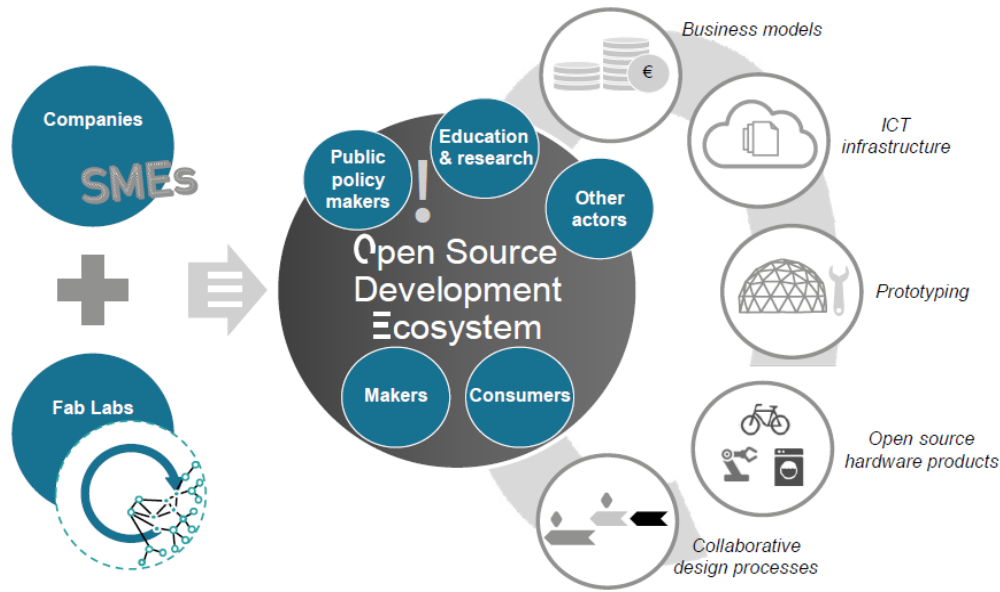

**Figure 6:** A concept by OPENNEXT [64]

## 5. Related work in open design movement

An inclusive review of variety of applications which are inspired by open design movement for developing smart applications is presented herein. Abid Abbasi et al. [65] developed a vibrant methodology for energy effectiveness using a small and cheap microcontroller which is centered on Solar Power Conditioning System using a high-level frequency inverter. Pulse width of the solar radiations is measured using high frequency invertors so as to get accurate and precise results. Hence, using such a system, an economical and energy efficient Power Conditioning System is developed for solar power applications. Alok Mukherjee et al. [66] presented a microcontroller-centered solar tracking system incorporating multiple functions for a solar cell is developed using PIC16F877A to monitor its output power. Thus, from this paper we can confidently conclude that an automatic sunlight adjusting system (ASAS) using solar power for the solar panel and control can be designed. It keeps solar panels directly to the sunlight. On reset the system will scan continuously in which direction the solar power is high. After correctly getting the position, the panel will follow the sunlight to get maximum power by switching over to the next quadrant. When the sun sets the solar panel will come to its reset position. D. Baskar et al. [67] developed an innovative method for preventing rail accidents. An AVR controller which works on the Keil C Software is used is for this rail transport application. It's primarily focus is to a system which can be used to prevent collisions between the trains and to aware the drivers about tracks condition for safety of everyone. The plan was to portray by using three distinct nodes and sensor duos in which databases of train are identified which are later used to give the evidence between the train and station with the help of serial port interaction. If there is a slight mismatch in the IP address, it will automatically alarm the driver who can then stop the train automatically or manually by using a buzzer. Future scope of this paper is to control the accidents and positioning the accidental Rail. Future prospect of this paper is to prevent rail accidents and ensure travel safety to its fullest. Dheepak Mohanraj et al. [68] proposed AVR centered control system which uses actual time data for logging temperature. This microcontroller is coded in Visual C++ language. Using a data logger, all accounts of the temperature controlling, and monitoring can be preserved. Data is logged & is transmitted by serial port to the computer. LCD screen displays output and the heater status. Hence, this system proves to be very efficient and gives accurate results. He presented an article where advancement of microcontroller centered speed control scheme of BLDC motor is carried out using the Proteus VSM Software. The controller used is PIC18F4331. The BLDC motor drive model was built for low power applications. The controlling method is censored wherein PIC18F4331 is the main controlling entity. Usage of this software greatly reduces the production time. Thus, the economy of this controller should possess business petition for low power application. Harshil Shah et al. [69] presented development and design of an economical microcontroller centered single phased water pump device. In order to achieve this, the voltage of the system is monitored using PIC12F675 controller. The software used was developed using the PCWH compiler. Hence, it is seen that this system was effectively developed and the execution was found to be very strong. This system employs a economical and local PIC12F675 microcontroller. To prevent the complicated A/D converter circuit and comparator, an A/D convertor and an internal comparator was utilized. Hence, the complete functioning of the system depends on the software which was developed. The design is simple and compact. In this proposal, the input ac voltage was varied from 198 volt to 248 volts, yet the output remained constant at 5-volt DC. The sensor which senses current along with its circuitry is designed such that, it can provide 2.5-volt dc to 5-volt dc. The presentation of the fully developed system was studied by equating its reliability cost and accuracy. It was found that a many large laboratory experiments ca be successfully conducted using this unit with acceptable outcomes. Last but not the least, it is the most economically developed system giving accurate results K. Lingeswaran et al. [70] developed a system in order to monitor one of the most important human body parameters- Blood Glucose. This is achieved using PIC 18F4520 controller which is programmed using the most basic and simple C Language. The outcomes are in accord with the benchmark blood glucose gauges. This paper demonstrates the use of a microcontroller for converting the glucose concentration in the sample of blood into voltages which are compatible with controller's input requirements. As soon as the sample is injected in the measuring device the blood glucose concentration is displayed on its screen. Using this open-sourced device, a fairly accurate and precise result was obtained. K. Govindaraju et al. [71] focused on the implementation and design of microcontroller centered automatic tracker for solar radiation. A sun tracking software is employed to supervise the power output and efficiency the of this tracker. It is used with a least number of parts and the usage of DC servo motors allows for the precise sun tracking. After scrutinizing the evidence, it is seen that this system can gather highest amount of energy instead of a conventional one. It is a feasible method for increasing the light energy. This method is easy, simple and requires no additional technical attention. This flexible software provides an efficiency of 21% over the static module. Thus, it is a very cheap device which has improved significant in solar tracking. M. Gopikrishnan et al. [72] presented design of autopilot of an unmanned aerial vehicle built on freescale Qorivva 32-bit microcontroller is carried out. Auto piloting is monitored using Freescale's MPC5644A controller which is programmed using its very own Freescale CodeWarrior software. For verification of the Autopilot software a mission plan comprising of six waypoints (home waypoint included) is successfully uploaded to the flight control computer. The aircraft successfully achieves all the waypoints and returns back to the Home waypoint in the HIL simulation. Constant monitoring of the system state is done using the GCS and visually through the FlightGear Flight simulator. Selection of the appropriate hardware platform and MEMS based sensors enables the system to achieve not only miniaturization and low power requirements but also higher accuracy for special mission requirements. The on-board processor is capable of achieving as high as 200Hz update rate for control loops making the system more accurate and suitable for a broad category of UAVs. V. V. Rangari et al. [73] developed an economical decay measurement system for Display Phosphors. This paper features the ATMEGA-16 controller programmed in the C and Assembly level language. A method measuring the decay time is designed successfully and satisfactory outcomes were obtained. This simple and economical system may be used to calculate the phosphor luminescence decay. However, as the time of measurement increases, the accuracy is found to decrease. Subitha M. B et al. [74] created a tracing system to find the need of Human Presence in Crucial Regions with the help of AT89C52 controller which is coded in C-Language. The growth of robots for local conditions is a tricky job. The rudimentary dilemma is exactly how to allow robots to identify and detect humans. Live human detection by PIR sensor is created by combining elements of all the hardware elements employed. Mohammad Arif Hossain et al. [75] developed of AVR Microcontroller based Home Automation System. The programming language used is the C-Language. Automation system is where home appliances can be remotely controlled and operated. This proven project used IR sensos to tally the attendance based on which the parameters were controlled and the total sum of people in the room were displayed. This system has low noise low power consumption and high efficiency. This system needs just 5V dc voltage for its operation. Priti K. Powale et al. [76] focused on the pressing need to deal with car theft which is done by developing an actual time antitheft system for automobile along with detection and prevention of accidents with the help of AVR Microcontroller which is coded in C-Language. The merit of this suggested system is that it can avoid the theft with the help of facial identification. If any approved individual enters into the vehicle then access will be given else the vehicle will get locked and its owner will be informed about the intrusion. Further, it can deliver password if the owner wants to give emergency access to someone whose face is not registered. Moreover, it can also carry out detection of accidents to provide travel safety by sending a message to owner and closest hospital and police station. Mohamed Abd El-Latif Mowad et al. [77] developed an android application using Arduino microcontroller for a smart home automated control system. This system functions on Bluetooth. In this paper the implementation and design of a monitor and control system has been established. This system comprises of many secondary systems which are controlled by microcontroller software. This system is also connected to a Bluetooth technology to control and monitor the electronic house components using both Arduino and micro controller from any place in the world. Bonvoisin & Mies [44] explains some projects undertaken are inspired from Open-Design Movement and which incorporate sustainability in their design. LifeTrac is an open-source tractor designed and created in the framework of the mission Open-Source Ecology. The product is intended to be simple and easy to gather to allow for DIY. This is accomplished by the usage of universal Lego-similar parts. Further, the product is intended to be modular, so it is easy to preserve, hence keeping its resilience. Multimachine is an open source many objective machine devices that is created to be manufactured by a non-adept employing frequently accessible tools from spare automobile parts. It is a 3-in-1 machine offering the purposes of lathe, milling, and drill press machines. This shall guarantee integration and upgradability of updated machine utilities using rotary elements. WikiHouse is an open-source home construction idea which is centred on devolved wooden structural modules. These can be locally manufactured by CNC machines and built by man without expert training. The project asserts to use sustainability-sourced timber and raw material with minimal fixed carbon. RepRap is a universal intent self-duplicating desktop 3D printer primarily employing the process Fuse Deposition Modelling. The engraved mechanical components employed in the composition of the machine are made of PLA, which is a thermoplastic that is appealed to be biodegradable and recyclable.

**6. Typical applications**

In this paper, a variety of applications are reviewed which are inspired by open design movement for developing smart applications. For further readings, many such systems have been presented such as Automation of small UAVs using a low cost MEMS sensor [78], novel digital PWM control [79], digital control technique for brushless DC motor [80], electronic white cane for blind people navigation assistance [81], survey on wearable obstacle avoidance electronic travel aids for blind persons [82], wireless security control system & sensor network for smoke & fire detection [83], face recognition technology [84], steering column locks and motor vehicle theft [85], a security module for car appliances [86], automobile anti-theft system design based on GSM [87], car alert system [88], autonomous mobile robotics [89] and condition monitoring related work [90-99]. Here, in figure 7, an open source and low-cost tools developed with support from federal funding are shown [113].

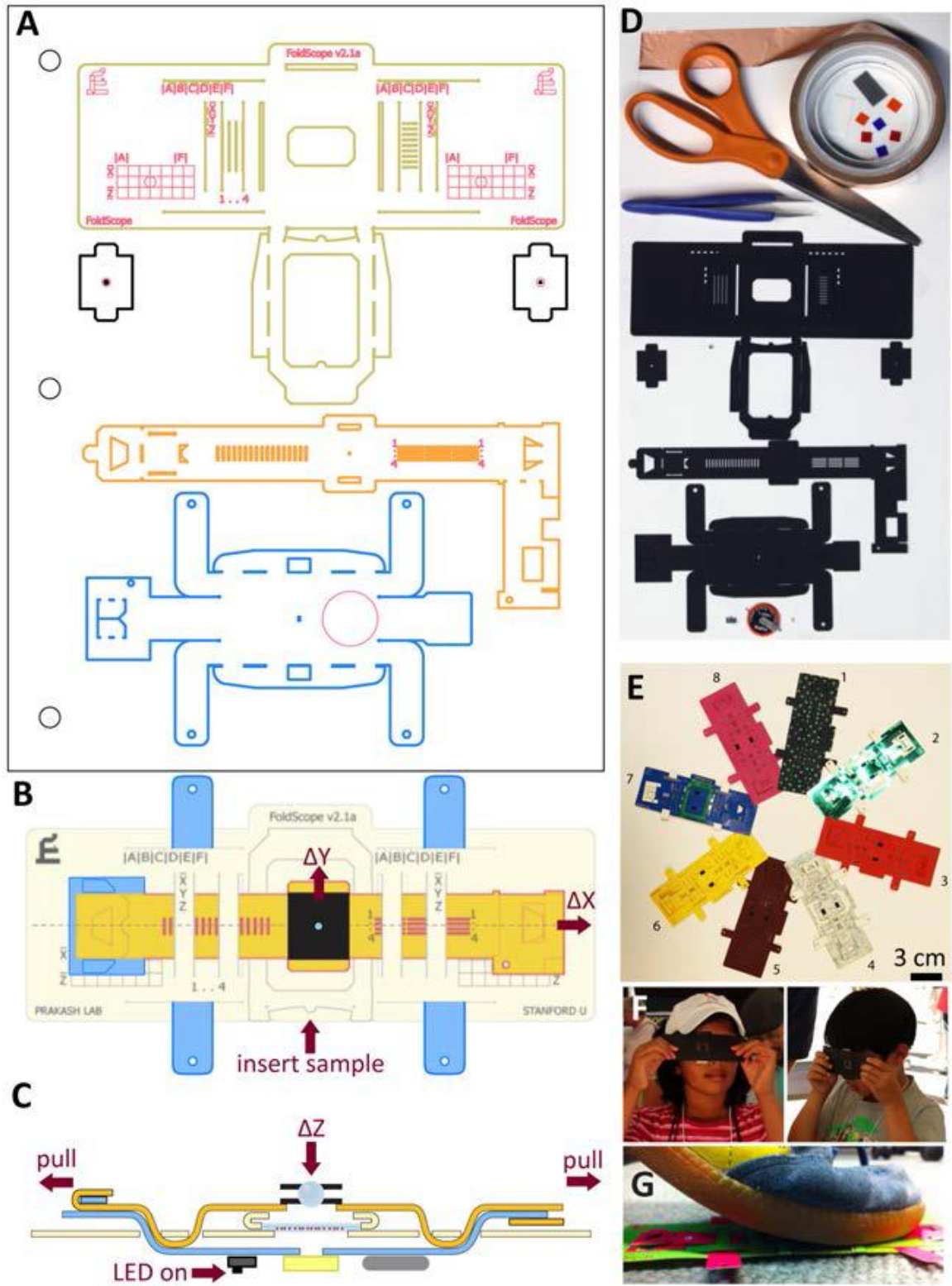

**Figure 7:** An open source and low-cost tools developed with support from federal funding [113]

Table 1 summaries all such applications with respect to software, parameter monitored, hardware, and sensors used.

**Table 1:** Summary table

| Application | Parameter Monitored | Controller Used | Software Used |
|---|---|---|---|
| Rail Transport | Collision Prevention | AVR | Keil C Software |
| Temperature Control System | Temperature | AVR | Visual C++ |
| Water Pump | Voltage | PIC12F675 | Compiler PCWH |
| Human Body | Blood Glucose | PIC 18F4520 | C Language |
| Solar Panel | Power Output, Efficiency | - | Sun Tracking Software |
| Unmanned Aircraft | Auto piloting | MPC5644A | Free scale CodeWarrior |
| BLDC Motor | Speed | PIC18F4331 | VSM Software |
| Display Instruments | Luminescence | Atmega-16 | C and Assembly Language |
| Humans | Human Presence | AT89C52 | C Language |
| Automobile | Accident Detection | AVR | C Language |
| Smart Home & Home Automation | - | Arduino AVR | Bluetooth |
| Robot | Motion | IR sensors | - |
| Photovoltaic System | MPPT Control | PIC16f877A | - |
| high power invertor | Protective system | - | - |

**7. Findings & implications**

The term open design is described as the directness of all associated records in a product development procedure, with the intention of cooperative development of substantial objects [100]. This is deemed as an instantiation of the notion of open-source innovation (OSI) [101], defined as the open revelation of data on a recent design with the intent of collective improvement of a one design or a reduced number of associated designs for non-market or market exploitation.

- An open-source design is stated as a prospective notion leading to an eco-efficient, complementary production and utilization of designs. Ecological design is a vital attempt to take up the disputes of sustainability and has principally focused on engineering design [102, 103]. One encouraging balancing method is the notion of Product Service Systems (PSS) – A combined deal of services and products concentrating on client delight instead on ownership of product [104]. This theory has been recognized as a 'possible sustainable corporate paradigm' since it may help end the connections between production and profit without restricting handling volume [105].

- An open-source design – A pioneering style is a kind of an open-source modernization, specified as the 'open uncovering of data on a different design with the purpose of collective advancement of a particular design or a restricted quantity of associated designs for market or non-market manipulation'.

| Why? **Advantages** & **Possibilities** to win with Open Source → | | Less Costs for R&D *Open Innovation, faster Bug Fixes* | Better Products *interactive, long-lasting, more Features, Freedom, Sustainability, Connectivity* | Less time to Market *Less legal Expenses, Less Time for Marketing* |
|---|---|---|---|---|
| Collaboration & Synergies *Material Cycles, Open Standards, Product-as-a-Platform, In-house Communication* | Ethical bonus for the Brand *Community, Sustainability, Education* | Less Costs for *Support, Ads, PR etc* | Better Employees | |
| | | Donations, Grants, Sponsoring, Public Research | Funding & Crowdfunding | Your Channels *Advertisement, Product-Partnerships, Rent, Fees* |
| Education & Training *Workshops, Certificates, Consulting, Events* | SUPPORT *Install, Operate, Maintain, Upgrade, Repair* | On Request *Individual Development, Customization, Adaptation* | Dual Licensing *Open only for non-commercial Use* | Closed Parts *Open Core, closed Add-ons, new Version closed, some Parts closed etc.* |
| Foundation/ Consortium Model *Members Fees* | Selling the Service *(e.g. Energy, Waste-Disposal, Food-growing, Data Collection) using OSHW* | Produce & sell Products *Quality, Warranted, Shipped* | ← How to make **Money** €/ Where does the Money come from? | |

**Figure 8:** Advantages and possibilities with open source [117]

- Numerous investigational manufacturing techniques assist public participants in manufacturing goods by themselves. Fabrication Laboratories (FabLabs) [106] are open and free engineering seminars delivering skilled animators and machines that permit individuals to understand their own developments of product innovation. On the other hand, manufacturing as a service (MaaS) [107] corporations can produce personalized goods on-demand and in small amount. Additional theories can encourage distributed and public-based fabrication. The CubeFactory [108], is a single cubic meter independent and portable construction structure incorporating four vital components: closed-loop physical supply, vigour supply, experience transfer, and manufacturing. Miniature manufacturing works are fabrication methods that can suit in a delivery vessel that can be shifted from a few places to an alternative to permit regional value concept [109]. Open design is therefore a major trend, and is sponsored by developments in manufacturing, knowledge, and association, and is seen as a hopeful return for any company. On the additional side, it challenges substantial questions that may be of significance for various technical specialties. How open or free a product is defined by three aspects such as availability, replicability, and clarity [110].

- The increase of ICT and inexpensive small-size production instruments are leading to the advancement of open design. This ground-breaking association of product development presents a fantastic prospect for uninterrupted development of products as well as a possibility for product improvement and growth of brand-new companies. In public-focused projects, all stages of the development are accomplished by an approximate throng of participants: expansion, production, product ideation, and prototyping. On the contrary, corporate-run projects are targeted at creating innovative results or developing current products to ultimately be produced and marketed by the corporation. Open design projects, since they are dependent on communications on the internet, additionally involve designed partnership platforms [111]. Figure 8 depicts advantages and possibilities with open source [117].

**8. Summary and future scope**

The comprehensive review carried here in which a variety of contemporary systems driven by open-design movement for smart applications are discussed. The products inspired from Open-design movement are the ones for which all the usage rights are without any restrictions granted to the common public and whose scientific credentials are entirely and freely accessible and available on the cyberspace. Though open-source and free-software has become broadly obtainable for everyone to use, alter, research, and circulate, products inspired from Open-design movement is an idea which still needs to go a long way before it becomes a widely used tool in various industries. This field has stemmed from the expansion of the open-source movement from software improvement into tangible physical products. For many centuries, businesses have devised and fabricated products using their own technological proficiency, which they then offer to the users. This procedure happens with the non-suitable trade-off that we as clients without even recognizing such a trade are bound to accept problems like planned obsolescence, redundant functions, and over-engineering. Then why resolve for the centrally monitored one-solution-for-all outcomes of closed modernization ecosystems? Suppose if someone was competent to tailor products of superior condition by using a crystal-clear source of knowledge and then get it made in their own locality for free. There exists need of projects that facilitates medium and small-sized organizations around the world to participate in societies with creators and users to profoundly transform the way the goods are produced, distributed, and designed. One such project is managed by Professor Roland Jochem [64] from the Technische Universität Berlin, and he gets together a group of collaborators from 19 EU nations, cementing the gap between consumer and industry. Figure 9 is a concept of open innovation ecosystem [112].

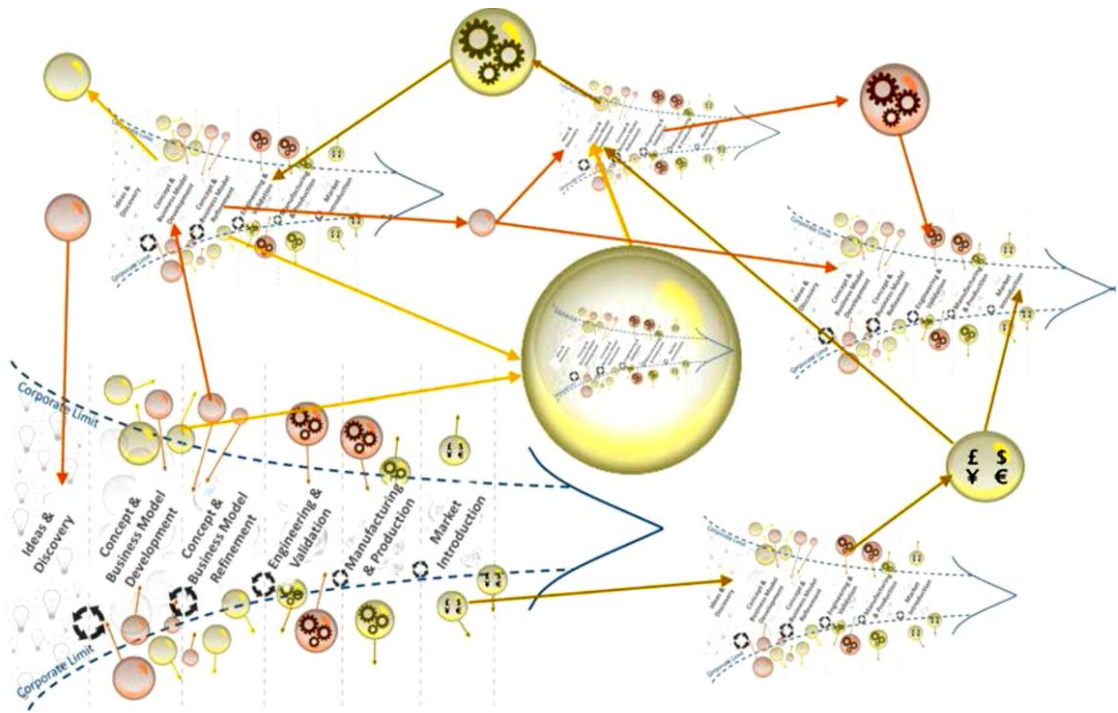

**Figure 9:** A concept of open innovation ecosystem [112]

The products inspired from open-design movement presents an immense promise for reforming the societal establishment of any development of product and restructuring the traditional engineering routine. There is need of the academics to boost perception of the mutual principles contained by the OSH territory and are proposing their assistance to the firms that are taking part in OSH societies.

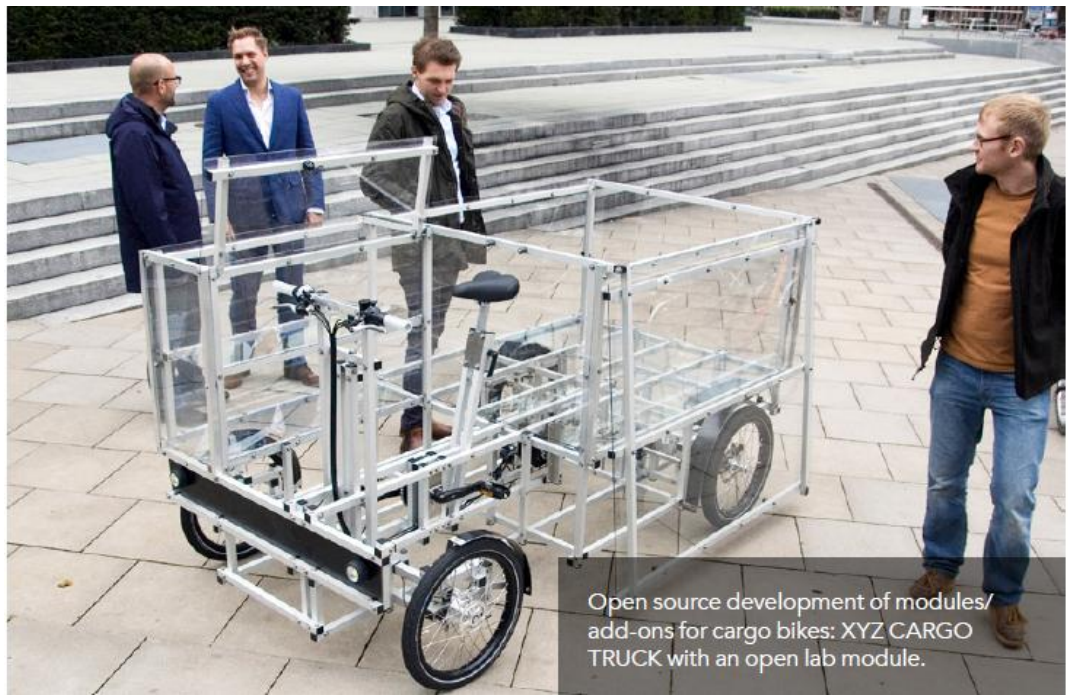

**Figure 10:** Open source development of modules/add-ons for cargo bikes: XYZ CARGO TRUCK with an open lab module [64]

The open-source style can possibly translate all production businesses. If buyers have free entry to particular methods and techniques of production, they can 3D print a replacement piece suitable for their product. Likewise, with restricted assets, SMEs will have the capability to modify results to meet their obligations. The products inspired from open-design movement arises alongside scenery of ever-increasing warmth to ecological and societal disputes which have navigated businesses on the road to incorporating brand new needs involving justice and eco-friendliness in their manufacturing endeavors to delight the consumers. Figure 10 is an open source development of modules/add-ons for cargo bikes: XYZ CARGO TRUCK with an open lab module [64]. Till today, the products inspired from open-design movement have been the realm of ordinary people, academic circles, and NGOs, and still hasn't reached a huge trade. If the products inspired from open-design movement is to become a majority occurrence, then conformance will become a crucial concern for both customers and fabricators. Instituting a place of such standards is believed to be challenging due to the multifactorial nature of employing the openness concept to bodily products. As a stride in the direction of founding a lucid OSH standard, the investigation squad built a device titled Open-o-Meter to evaluate the candidness of substantial merchandise. This device makes its way for evaluating process and product openness and further enables the client to verify if technical information of a product will let anybody to distribute it. It delivers a straightforward list for its representatives to decide the hard work of originator of product to meet the tenets of open source. Further, it gives its specialists with procedures for supervising the data throughout the growth. Additionally, it uncovers the contextual and multifunctional characteristics of honesty and gives a starting point for the scholars to scrutinize if the principles of sincerity are not obligatory or compulsory in a standardization move.

- This movement is at present nascent, yet it holds enormous potential for the years down the lane. In some ways engineering and design are further acclimatized to mutual progress instead of the progressively generic open-sourced software system assignments.
- It is not compulsory that the partners should communicate in the same language to effectively work in partnership. Nevertheless, there are obstacles to conquer for this design to come into existence as equated to development of software which includes more sophisticated and commonly used instruments available.
- Creating, modifying and testing designs which are physically developed is not direct since time, cost and effort are vital to generate a physical piece; though having access to evolving elastic automated manufacturing procedures the difficulty and struggle of building can be greatly reduced.